# Mechanism of Tulip Flame Formation in Highly Reactive and Low Reactive Gas Mixtures

Chengeng Qian (钱琛庚)[a] and Mikhail A. Liberman [b*]

[a] *Aviation Key Laboratory of Science and Technology on High Speed and High Reynolds Number, Shenyang Key Laboratory of Computational Fluid Dynamics, Aerodynamic Force Research AVIC Aerodynamics Research Institute,*
*Shenyang 110034, China*

[b] *Nordita, KTH Royal Institute of Technology and Stockholm University, Hannes Alfvéns väg 12, 114 21 Stockholm, Sweden*

**Abstract**

The early stages of flame dynamics and the development and evolution of tulip flames in closed tubes of various aspect ratios and in a half-open tube are studied by solving the fully compressible reactive Navier–Stokes equations using a high-order numerical method coupled to detailed chemical models in stoichiometric hydrogen/air and methane/air mixtures. The use of adaptive mesh refinement (AMR) provides adequate resolution of the flame reaction zone, pressure waves, and flame-pressure wave interactions. The purpose of this study is to gain a deeper insight into the influence of chemical kinetics on the combustion regimes leading to the formation of a tulip flame and its subsequent evolution. The simulations highlight the effect of flame thickness, flame velocity, and reaction order on the intensity of the rarefaction wave generated by the flame during the deceleration phase, which is the principal physical mechanism of tulip flame formation. The obtained results explain most of the experimentally observed features of tulip flame formation, e.g. faster tulip flame formation with deeper tulip shape for faster flames compared to slower flames.

Emails:
Chengeng Qian: qiancg@avic.com
[*]Mikhail A. Liberman (corresponding author): mliber@nordita.org

# 1. Introduction

Burning fossil fuels is believed to significantly increase the amount of carbon dioxide in the atmosphere, leading to climate change and global warming. A natural way to re-duce carbon emissions and avoid these problems is to produce clean energy using hydro-gen. However, easy leakage and high flammability make hydrogen application difficult due to high risk of explosion [1]. A possible solution to this problem, which has been intensively studied experimentally and numerically in recent years, is the use of carbon or ammonia fuels mixed with hydrogen. [2-5]. The injection of hydrogen into natural gas is an effective way to improve combustion in the power industry. Blending hydrogen with natural gas can reduce carbon emissions and increase energy efficiency. Hydrogen additive can improve the combustion of hydrocarbon fuel due to its low ignition energy, high reactivity and fast burning rate, making hydrogen enriched natural gas a promising alternative fuel. Depending on the percentage of hydrogen added to methane or ammonia, the ignition delay time can be significantly reduced, which is important for its industrial application, such as direct injection engines [6] or diesel engines [7].

Therefore, there are currently extensive experimental and numerical studies on the combustion of premixed methane/hydrogen/air mixtures in closed or semi-open tubes to gain a deeper insight into the different combustion regimes and efficiencies, see e.g. [5-8]. The effect of laminar flame velocity on flame dynamics was investigated by Deng et al. [9] using a two-dimensional laminar combustion model and simplified reaction mechanisms derived from the one-step methane/air reaction model. The effect of reaction order on flame propagation was studied by Qi et al [10] using numerical models with three single step reaction mechanisms .

Flame propagation in tubes is important for understanding combustion processes under confined conditions, such as explosions and safety issues, as well as for industrial and technological applications. It provides a fundamental physical-chemical platform for the analysis of more complex engineering and process problems and for the development of appropriate analytical methods and numerical models. Therefore, understanding the structure and acceleration of premixed flames is of primary interest to the combustion community.

It is well known that a flame ignited near the closed end of a tube and propagating to the opposite closed or open end suddenly slows down, and the shape of the flame front rap-idly changes from a convex with the tip pointing forward to a concave shape with the tip pointing backward. This phenomenon has been first photographed in experiments by Ellis [11] and later named "Tulip Flame" (TF). Subsequent experimental studies have shown that tulip flame formation is remarkably robust across combustible mixtures, highly or slowly reactive mixtures, and downstream conditions: such as open or closed tube ends and tube shapes, tube wall roughness, and obstructed tubes [12, 13]. The numerical study [14] has shown that the formation of the tulip flame is similar for adiabatic and isothermal wall boundary conditions. It has also been shown in [15, 16] that the geometry of the ignition source is not critical for flame dynamics and tulip flame formation. The transition to a tulip flame always occurs in tubes with a sufficiently large aspect ratio $L/D > 4$ for both laminar and turbulent flames. Only with relatively strong breaking of the initial symmetry, e.g. with sufficiently strong perturbations of the initial flame front, tulip flame formation does not occur. Thus, the formation of tulip flames can be considered as one of the most fundamental phenomena of combustion physics.

It should be noted that flame front inversion can be caused by various processes, such as flame collision with a shock wave or pressure wave, various hydrodynamic instabilities inherent to the propagating flame, such as Darrieus-Landau (DL) or Rayleigh-Taylor (RT) instabilities. This makes the concept of tulip flame formation somewhat ambiguous, which has led to a large number of different scenarios that have been explored in an attempt to explain the physical mechanism of tulip flame formation. Flame collision with a shock wave [17], flame front instabilities such as DL instability [18] and RT instability [19, 20] have been considered as possible mechanisms of tulip flame formation. Flame-vortex interaction has been considered by many researchers [21 - 25] as a seemingly plausible scenario for flame front inversion and tulip flame formation, although no evidence has been provided that this is actually the case. The reader is referred to the reviews [12, 13] for references to numerous studies of the physical mechanisms associated with tulip flame formation. However, numerous theoretical, computational, and experimental studies have failed to provide a convincing and experimentally consistent explanation for the physical mechanism of tulip flame formation, and until recently this problem remained one of the fundamental unsolved problems in combustion science.

The tulip flame considered in the present study is the rapid inversion of the flame front from the convex to the concave shape with the flame tip pointing toward the burned combustion products, which occurs during the deceleration phase associated with the reduction of the flame surface area when the lateral parts of the finger-shaped flame touch the tube sidewalls. Recently, Liberman et. al [26] have shown for the first time that the rarefaction wave generated by the flame during the deceleration phase is the key process in tulip flame formation. This mechanism of the tulip flame formation explains various phenomena observed in experimental studies and numerical simulations, such as reverse flow, adverse pressure gradient, etc. This also means that tulip flame formation is a purely hydrodynamic phenomenon, as it occurs much faster than the time required for flame front instabilities to develop, which is consistent with the experimental study of tulip flame formation by Ponizy et al. [27].

Recent experimental and numerical studies [6-8, 28-30] have revived interest in the physical mechanism of tulip flame formation at different blending of methane and hydrogen mixtures. The purpose of the present work is to gain a deeper insight into the influence of chemical kinetics on the combustion regimes leading to the formation of a tulip flame and its subsequent evolution for a flame in the highly reactive $H_2$/air mixture compared to a flame in the slowly reactive $CH_4$/air mixture. To do this, we compare and contrast unsteady, fully compressible high accuracy solutions of the Navier–Stokes equations coupled with the multi-step detailed chemical models for hydrogen-air flames with the solutions for methane-air flames. Based on these two limiting cases: pure $H_2$/air flame and pure $CH_4$/air flame, it is easy to understand the hydrogen/methane/air flame dynamics for different percentages of hydrogen admixture.

The paper is organized as follows. In Section 2 the details of numerical simulations, chemical models and boundary conditions are presented. In Section 3.1 – 3.3 we present results from a series of two-dimensional simulations of $H_2$/air and $CH_4$/air flames propagation in closed tubes of various aspect ratios and in a semi-open tube. The results of these simulations are analyzed and compared with experimental studies of $H_2$/air and $CH_4$/air flames propagation in closed and semi-open tubes. In the last section "Summary and Discussion" we show that the tulip formation due to the rarefaction wave is much faster than the characteristic times of the flame instabilities and summarize the results obtained in the present study.

## 2. Numerical models

### 2.1. Two-dimensional DNS

Our previous studies [16, 26] have shown that the flame dynamics obtained from 2D modeling are qualitatively very similar to those observed in experimental studies. However, it should be emphasized that since the acceleration/deceleration processes in 3D flames are much faster than in 2D flames, 2D modeling provides good qualitative but not quantitative agreement with experimental data. In the present work the two-dimensional computational domains are modelled using high-resolution DNS of the fully compressible reactive Navier–Stokes equations coupled to detailed multistep chemical models for combustion of a stoichiometric hydrogen/air and methane/air mixtures. A detailed chemical kinetic model for a stoichiometric hydrogen/air mixture consisting of 21 reactions and 9 species developed by Li et al [31] was implemented and used in the simulations. The multistep detailed chemistry for a stoichiometric methane/air is the reduced de-tailed mechanism 15S-26R in [32], which consists of 15 species and 26 reversible reaction steps. This reduced mechanism has been calibrated by freely propagating laminar flame with adiabatic boundary. The laminar flame velocities, mass fractions of target species, heat re-lease obtained by 15S-26R mechanism agree well with those obtained by more complex detailed chemical model DRM-19 [33] and GRI 1.2 [34]. In our previous paper [36], we have shown that the results obtained by using DRM-19 is very closed to those by using GRI 3.0 [35] and experimental data. So that the reduced mechanism 15S-26R is considered as smallest chemical model which can produce correct important properties of flame, such as laminar flame velocities, mass fraction of species, et.al. and is employed in the simulations of CH4/air in this paper. The considered 2D computational domains are rectangular channels of width $D = 1 \text{cm}$ with both ends closed and aspect ratios of $L/D = 6, 12$ and a rectangular channel of width $D = 1 \text{cm}$ with an open right end. Formally, a semi-open channel can be thought of as a channel with an infinitely large aspect ratio. From a physical point of view, the main feature of a semi-open channel is the absence of reflected pressure waves. In Section 3.3, we will show that without reflected pressure waves, flame shape evolution and tulip flame formation in a semi-open channel takes longer and proceeds much more smoothly than in a channel with both ends closed.

The numerical simulations solve the 2D time-dependent, reactive compressible Navier-Stokes equations including molecular diffusion, thermal conduction, viscosity and detailed chemical kinetics. The governing equations are

$$\frac{\partial \rho}{\partial t} + \frac{\partial (\rho u_i)}{\partial x_i} = 0, \tag{1}$$

$$\frac{\partial (\rho u_i)}{\partial t} + \frac{\partial (P\delta_{ij} + \rho u_i^2)}{\partial x_j} = \frac{\partial \sigma_{ij}}{\partial x_j}, \tag{2}$$

$$\frac{\partial (\rho E)}{\partial t} + \frac{\partial \left[(\rho E + P)u_i\right]}{\partial x_i} = \frac{\partial (\sigma_{ij} u_i)}{\partial x_i} - \frac{\partial q_i}{\partial x_i}, \tag{3}$$

$$\frac{\partial \rho Y_k}{\partial t} + \frac{\partial \rho u_i Y_k}{\partial x_i} = \frac{\partial}{\partial x_i}\left(\rho V_{ik} Y_k\right) + \dot{\omega}_k, \tag{4}$$

$$P = \rho R_B T \left(\sum_{i=1}^{N_s} \frac{Y_i}{W_i}\right), \tag{5}$$

$$\sigma_{ij} = \frac{4}{3}\mu\left(\frac{\partial u_i}{\partial x_j} + \frac{\partial u_j}{\partial x_i}\right) - \frac{2}{3}\mu \frac{\partial u_i}{\partial x_i}\delta_{ij}, \tag{6}$$

$$q_i = -\kappa \frac{\partial T}{\partial x_i}. \tag{7}$$

The quantities $\rho$, $u_i$, $T$, $P$, $E$, $\sigma_{ij}$, $q_i$ are density, velocity components, temperature, pressure, specific total energy, viscosity stress tensor, and the heat fluxes. $R_B$ is the universal gas constant, $Y_i, W_i, V_{i,j}$ are mass fraction, molar mass and diffusion velocity of species $i$. The viscosity, thermal conduction and diffusion coefficients of the mixture are calculated by the method presented in [37]. The reaction rate of species $k$ is determined as

$$\dot{\omega}_k = W_k \sum_{j=1}^{N_r} (v''_{jk} - v'_{jk}) \cdot \left(k_{f,j} \prod_{s=1}^{N_s} \left(\frac{\rho Y_s}{W_s}\right)^{v'_{js}} - k_{b,j} \prod_{s=1}^{N_s} \left(\frac{\rho Y_s}{W_s}\right)^{v''_{js}}\right), \tag{8}$$

where $v'_{js}$ and $v''_{js}$ are stoichiometric coefficients of species $k$ of the reactant and product sides of reaction $j$. $k_{f,j}$ and $k_{b,j}$ are the forward and backward reaction rates of reaction j, which are calculated by Arrhenius equation.

The governing equations are solved using open-source code AMReX-Combustion PeleC [38], which is a load-balanced Adaptive Mesh Refinement (AMR) compressible reacting flow solver. The convection terms are discretized by classical 5th order WENO-JS scheme of Jiang and Shu [39]. The 2nd order centered difference method is used to calculate the diffusion terms. The time integral method is based on a spectral deferred correction approach [40]. The integrator of chemical source terms is the explicit Runge-Kutta methods in the ARKODE packages of SUNDIALS [41].

An adaptive mesh refinement (AMR) method was used to reduce the computational time required for the numerical simulations with detailed chemistry for $H_2$/air and $CH_4$/air flames. The criterion for grid adaptation was a mass fraction of $HO_2$ in the computational domain. The AMR method in PeleC is based on the AMReX framework [42], which supports massively parallel, block-structured adaptive mesh refinement. See **Appendix A** for a more detailed discussion of grid resolution and convergence. The non-reflecting outflow boundary condition [43] was used to calculate the subsonic outflow.

Reliable modeling of reactive flows requires adequate resolution of the internal structure of the flame. Since the thickness of a laminar flame decreases with increasing pressure, higher resolution is required to determine the flame structure at elevated pressures. For hydrogen/air flame the reaction order is $n=2$, which means that the hydrogen/air flame does not depend on pressure since $U_f \propto P^{\frac{n}{2}-1}$ [44]. The method used to calculate the reaction order is described in **Appendix B**. For the flame thickness we obtain

$$L_f \approx \nu / U_f \propto P^{-n/2}, \tag{9}$$

where $\nu$ is kinematic viscosity. Therefore, the flame thickness decreases with pressure as $L_f \propto 1/P$. The reaction order of methane/air flame is $n \approx 1.1$. In the same way for the methane/air flame we obtain, $U_f \propto 1/P^{1/2}$ and $L_f \propto 1/P^{0.55}$. The maximum pressure during tulip flame formation in a channel with closed ends and aspect ratio $\alpha = 6$ is about 2atm for a hydrogen/air flame, resulting in a reduction in flame front thickness from $L_f = 350\,\mu m$ to $L_f \simeq 100\,\mu m$.

The uniform coarse grid resolution of 40μm was used. With each refinement, the grid size is halved, allowing for a maximum of two refinements. As a result, the grid resolution near the flame front reaches 10μm corresponding to 10 grid points per flame front, which is sufficient to predict the correct laminar flame structure and the flame velocity. Thorough resolution and convergence (grid independence) tests have been performed in our previous publications [26, 16] by varying the value of the computational grid size $\Delta x$ to ensure that the resolution is sufficient to capture details of the problem in question and to avoid computational artifacts. The reader is referred to [16, 26] for details of the convergence and resolution tests.

### 2.2. Boundary and initial conditions; modeling parameters

The flame is ignited by a small (3mm) semi-circular pocket of burned gas at adiabatic flame temperature, $T_b = 2350K$ for $H_2$/air and $T_b = 2231K$ for $CH_4$/air, near the left closed end. The initial conditions are $P_0 = 1\,atm$, $T_0 = 298\,K$. The center of the ignition spot is on the mid-plane of the channel near the left closed end. The adiabatic, no-slip reflective boundary conditions at the channel walls are

$$u_x = 0, \ \partial T / \partial \vec{n} = \partial Y_k / \partial \vec{n} = 0, \tag{10}$$

where $\vec{n}$ is the normal to the wall. The $x$ coordinate is taken along the channel walls, $y = 0$ is the mid-plane coordinate and the sidewalls are at $y = \pm D/2$.

During tulip flame formation, symmetry boundary conditions can be applied at the mid-plane, $y = 0$, between the plane sidewalls

$$\left.\frac{\partial \vec{u}}{\partial y}\right|_{y=0} = 0, \ \left.\frac{\partial T}{\partial y}\right|_{y=0} = 0, \ \left.\frac{\partial Y_k}{\partial y}\right|_{y=0} = 0. \tag{11}$$

The symmetric formulation of the problem is justified by the fact that symmetric tulip flame formation relative to the tube axis has been observed in many experimental studies, see e.g. [6, 8, 24]. Depending on the experimental conditions, the gravitational buoyancy effect can cause asymmetric shape of the tulip flame. However, the characteristic time for the buoyancy is usually larger than the time of a tulip flame formation even for methane/air flames.

The model parameters used in simulations are shown in Table 1 for hydrogen/air mixture, and in Table 2 for methane/air mixture.

**Table 1.** Model parameters for simulating a stoichiometric hydrogen–air flame.

| Initial pressure | $P_0$ | 1.0 atm |
|---|---|---|
| Initial temperature | $T_0$ | 298 K |
| Initial density | $\rho_0$ | $8.5 \cdot 10^{-4}$ g/cm$^3$ |
| Laminar flame velocity | $U_f$ | 2.43 m/s |
| Laminar flame thickness | $L_f$ | 0.0325 cm |
| Adiabatic flame temperature | $T_b$ | 2503 K |
| Expansion coefficient ($\rho_u / \rho_b$) | $\Theta$ | 8.34 |
| Specific heat ratio | $\gamma = C_P / C_V$ | 1.399 |
| Sound speed | $a_s$ | 408.77 m/s |

**Table 2.** Model parameters for simulating a stoichiometric methane–air flame.

| Initial pressure | $P_0$ | 1.0 atm |
|---|---|---|
| Initial temperature | $T_0$ | 298 K |
| Initial density | $\rho_0$ | $1.1\times 10^{-3}$ g/cm$^3$ |
| Laminar flame velocity | $U_f$ | 0.3803 m/s |
| Laminar flame thickness | $L_f$ | 451 μm |
| Adiabatic flame temperature | $T_b$ | 2231 K |
| Expansion coefficient ($\rho_u/\rho_b$) | $\Theta$ | 7.2 |
| Specific heat ratio | $\gamma = C_P/C_V$ | 1.38 |
| Sound speed | $a_s$ | 332 m/s |

## 3. Results

In Sections 3.1-3.3, we use 2D simulations to compare the flame dynamics leading to the flame front inversion and tulip flame formation in a highly reactive H2/air mixture and a slowly reactive CH4/air mixture in channels of different aspect ratios and in a half-open channel. In previous study [26] it was shown the rarefaction wave generated by the flame during the flame deceleration phase is the primary physical process leading to flame front inversion and tulip flame formation. The underlying physics of the flame front inversion is very simple if we consider a theoretical thin flame model [44, 45], where the flame is considered as a discontinuous surface separating unburned and burned gases. During the acceleration phase, the flame acts as a semi-transparent accelerating piston, generating pressure waves whose amplitude is greater the greater the acceleration of the flame (piston). The initial rarefaction wave is generated by the flame as the flame skirt touches the sidewalls of the tube, resulting in a reduction in flame surface area and a subsequent reduction in flame velocity. In the reference frame of the unburned flow, the decelerating flame is similar to the piston that begins to move with acceleration out of the tube (to the left). It is well known [46] that such a piston generates a simple rarefaction wave, whose "head" propagates forward (to the right) with the sound speed, creating a reverse flow of the unburned gas between the piston and the head of the rarefaction wave. The velocity of the unburned gas in the reverse flow is maximum (equal to the piston velocity) near the piston surface and tends to zero toward the head of the rarefaction wave. It is important to note that in the classical problem of a rarefaction wave initiated by a flat piston (one-dimensional problem) [46], the reverse flow velocity is uniform. On the contrary, the reverse flow velocity generated by a convex piston is not uniform. Near the piston, the reverse flow axial velocity is maximum at the tube axis and decreases toward the side walls. The resulting flow in the unburned gas is a superposition of the unburned gas flow created by the flame during the acceleration phases and the reverse flow created by the rarefaction wave. It is obvious that in the immediate vicinity of the flame front, the unburned gas flow acquires a mirror tulip-shaped profile of axial velocity. In the thin flame model, the velocity of a small portion of the flame front is equal to the sum of the laminar flame velocity at which the flame propagates relative to the unburned gas and the velocity of the unburned gas immediately ahead of that portion of the flame front at which the unburned gas entrains that portion of the flame front. Therefore, the mirror tulip velocity profile formed in the unburned gas in the close vicinity ahead of the flame leads to inversion of the flame front and formation of a tulip flame. In [26], the results of the thin flame model were found to be in surprisingly good agreement with 2-D and 3-D simulations. In closed tubes, the collision of a flame with a pressure wave reflected from the opposite end of the tube will also cause the flame to decelerate, thereby increasing the effect of the first rarefaction wave or leading to a distorted tulip flame if the deceleration caused by the flame collisions with pressure waves is sufficient for the RT instability to develop.

**3.1 2D channel with both ends closed and aspect ratio $\alpha = L/D = 6$**

In this section we compare the results of modelling dynamics of hydrogen/air flame with dynamics of methane/air flame in a 2D channel with both ends closed and aspect ratio $\alpha = 6$. Fig. 1(a) shows the time evolution of the flame front velocities along the channel axis, $U_f(y=0)$, and the flame velocity close to sidewall, $U_f(y=0.4cm)$, calculated for hydrogen/air flame. Fig. 1(b) shows the calculated time evolution of the flame front velocities along the tube axis, $U_f(y=0)$ and near the side wall, $U_f(y=0.38cm)$, calculated for methane/air flame. Both figures also show the pressure increase $P_+$ and the pressure growth rate $dP_+/dt$ at the centerline just ahead of the flame front. The pressure growth rate shows the location of the reflected pressure waves immediately before their collisions with the flame.

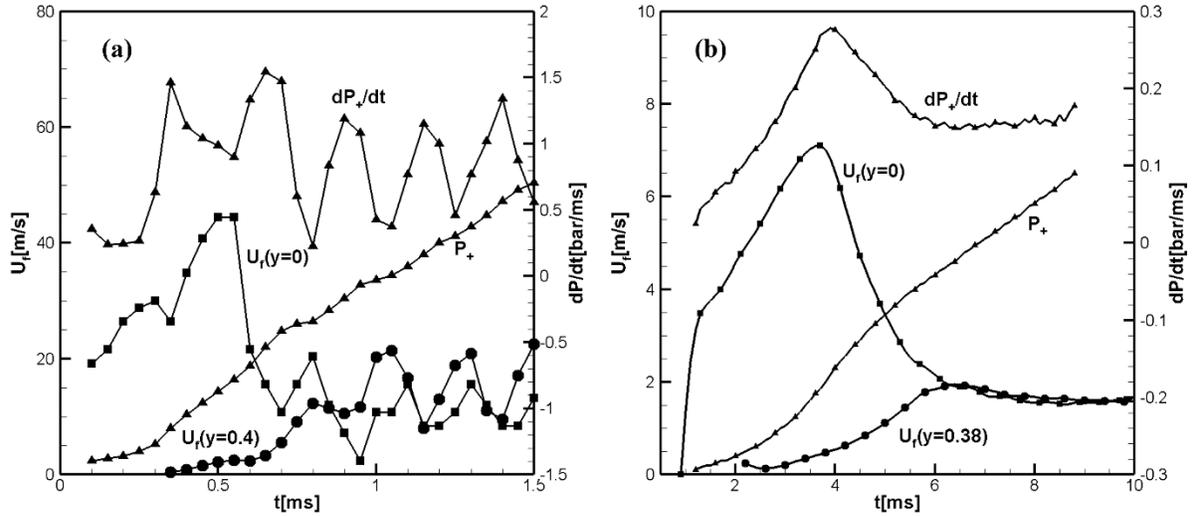

**Figure 1**: (a) Temporal evolution of the $H_2$/air flame front velocities at $y=0$ and at $y=0.4cm$; (b) $CH_4$/air flame front velocities at $y=0$ and at $y=0.38cm$.

Figure 2 shows: (a) temporal evolution of the unburned flow velocities at the tube axis, $u_+(y=0)$ and near the sidewall, at $y=0.4cm$, calculated closely ahead of the flame front, at $Y_{HO_2} < e^{-8}$, for hydrogen/air flame; and (b) similar values $u_+(y=0)$ and $u_+(y=0.38cm)$ for $CH_4$/air flame. Dashed lines in Fig. 2(a, b) show the differences $\Delta u_+ = u_+(y=0.4cm) - u_+(y=0)$ in Fig. 2a and $\Delta u_+ = u_+(y=0.38cm) - u_+(y=0)$ in Fig. 2b. It can be seen that for hydrogen/air the velocity in the unburned gas near the sidewall exceeds the unburned gas velocity at the channel axis (mid-plane), $\Delta u_+ > 0$ at $t \approx 0.57 ms$. Shortly thereafter, the axial velocity profile in the unburned gas begins to take on a tulip-shape, which leads to the inversion of the flame front and the formation of a tulip shape flame, as can be seen in Fig. 3a.

In general, the tulip flame formation scenario in the methane/air shown in Fig. 2b is similar to that for the hydrogen/air flame in Fig. 2a. The main difference is that the tulip flame formation for hydrogen/air flame occurs about ten times faster than for methane/air flame due to the much larger value of the negative acceleration of the hydrogen-air flame during the first deceleration phase, which causes a much more intense initial rarefaction wave and, consequently, a much faster formation of the tulip flame. Another difference is the much greater acceleration of the hydrogen-air flame during the finger stage, just before the decelerating stage. Therefore, during this stage the accelerating hydrogen-air flame generates strong pressure waves, which can be seen as the pressure growth rate $dP_+/dt$ in Fig. 1a. The amplitude of the pressure waves generated by accelerating hydrogen/air flame is much greater than the amplitude of pressure waves generated by methane/air flame.

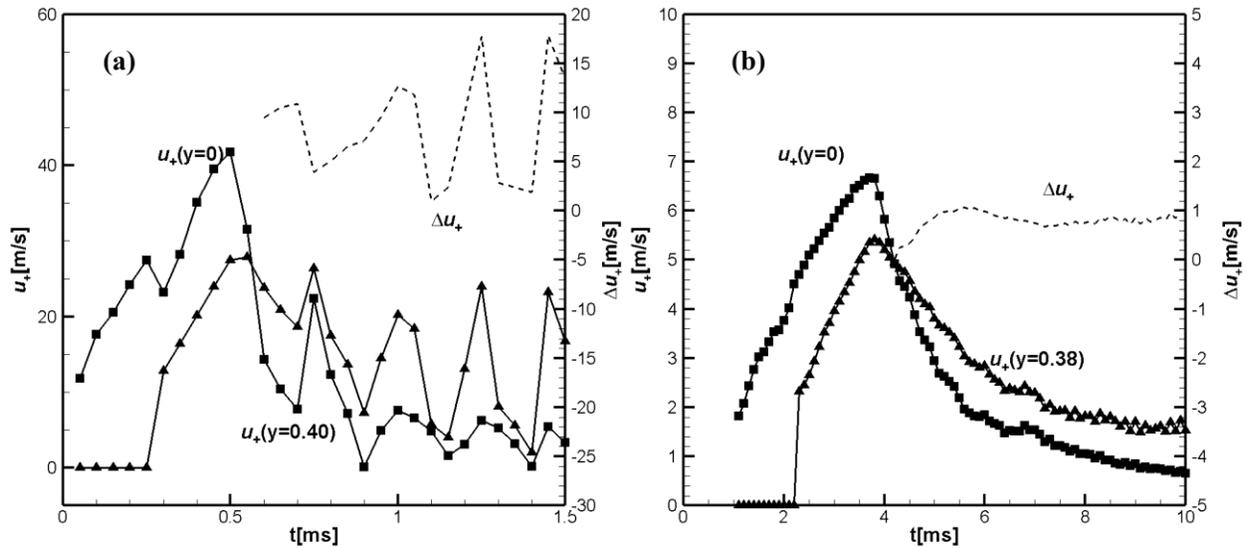

**Figure 2**: (a) Temporal evolution of the unburned flow velocities $u_+(y=0)$ and $u_+(y=0.4cm)$ in H$_2$/air; and (b) in CH$_4$/air $u_+(y=0)$ and $u_+(y=0.38cm)$.

The strong pressure waves reflect from the opposite closed end of the channel travelling back and forth. They first enhance the effect of the initial rarefaction wave. Later, their collisions with the tulip flame slowdown the flame, and can cause a sufficiently large negative acceleration of the flame to induce the Rayleigh-Taylor instability leading to the formation of a distorted tulip flame [24, 25]. This can be seen in Fig. 3a at $t=1.3\,\text{ms}$. On the contrary, in case of a methane/air flame the pressure waves are very weak and do not cause the flame velocity oscillations seen in Fig.1a and Fig.2a, or a flame instability, which can be seen in Fig. 3a.

Figure 3 shows computed schlieren images and streamlines for H2/air flame (a); and for CH4/air flame (b) for selected times during tulip flame formation. The red dashed lines in the figures show the location of the unburned gas axial velocity profile at 0.5mm ahead of the flame front, where it was measured.

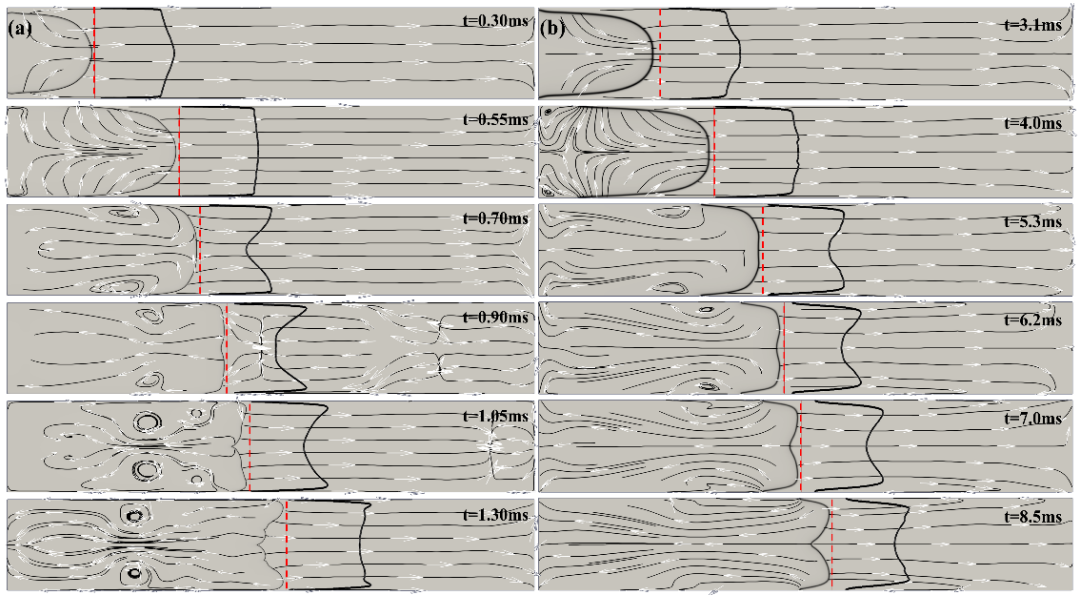

**Figure 3:** (a) Time sequence of computed schlieren images and streamlines for H$_2$/air; (b) for CH$_4$/air flame.

It can be shown that the magnitude of flame acceleration/deceleration is strongly de-pendent on flame velocity. The greater the flame velocity, the greater the magnitudes of flame acceleration and deceleration. Since the intensity of the rarefaction wave generated by the decelerating flame is characterized by the magnitude of the flame deceleration, a faster flame will produce a stronger rarefaction wave. Therefore, for a faster flame, the difference between the unburned gas velocity at the tube axis and the axial (along the tube) velocity closer to the sidewalls is greater than for a slower velocity flame. This means that a higher velocity flame will transition to a tulip flame with a deeper tulip (longer tulip petals) and more quickly than a lower velocity flame. This can be clearly seen by comparing Fig. 3(a) and 3(b). On the contrary, in the case of a methane/air flame, the pressure waves are very weak and do not cause the flame velocity oscillations seen in Fig. 1a and Fig. 2a, or the flame instability seen in Fig. 3a. Fig. 3b shows that the methane/air flame is formed smoothly, without oscillations or instabilities.

**3.2. 2D channel with both ends closed and aspect ratio $\alpha = 12$**

Figure 4 shows the temporal evolution of the flame front velocities along the mid-plane of the channel and near the sidewall, the pressure rise, $P_+$, and the pressure growth rate $dP_+/dt$ at the centerline ahead of the flame front calculated for the hydrogen/air (Fig.4a) and methane/air (Fig. 4b) flames propagating in a 2D channel with aspect ratio $\alpha = 12$.

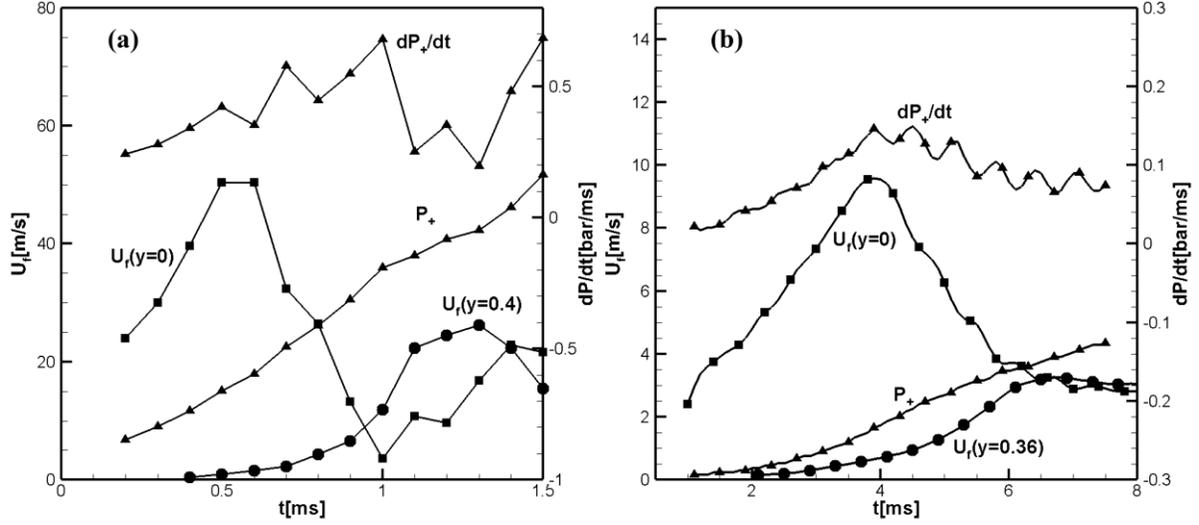

**Figure 4**: (a) Temporal evolution of velocities $U_f(y)$ on the flame front at $y=0$ and $y=0.4 cm$ for H$_2$/air flame; (b) the same for CH$_4$/air flame for $y=0$ and $y=0.36 cm$.

It can be seen in Fig.4a that there are fewer reflected pressure waves colliding with the flame compared to the shorter channel with aspect ratio $\alpha=6$ shown in Fig.1a, where it can be seen that the collision of the flame with the reflected pressure wave at 0.25ms results in a lower maximum flame front velocity $U_f(y=0)=45 m/s$, while in Fig. 4a the maximum flame front velocity along the center line is $U_f(y=0)=50 m/s$. A comparison of Fig. 3a and Fig. 6a shows that in the longer channel, flame collisions with reflected pressure waves, if they can lead to the formation of a distorted tulip flame, do so at a later time than in a shorter channel. The laminar velocity of methane/air flame is small and therefore the scenario of tulip flame formation is practically the same for channels with aspect ratio $\alpha=6$ and $\alpha=12$.

Figure 5 shows: 5a the temporal evolution of the unburned gas velocities at 0.5mm ahead of the flame front for hydrogen/air, and 5b the same for the methane/air flame. A higher unburned flow velocity along the centerline $u_+(y=0)$ compared to that in the shorter channel shown in Fig.1(a) is the result of a higher flame front velocity $U_f(y=0)$. Due to the reduced number of collisions with the reflected pressure wave, the distorted tulip flame may develop at a later time, if at all. Except for small fluctuations in unburned flow velocities $u_+(y=0)$ and $u_+(y=0.36cm)$ caused by weak pressure waves, the whole scenario of tulip flame formation in $CH_4$/air is very similar to that in the shorter channel $\alpha=6$.

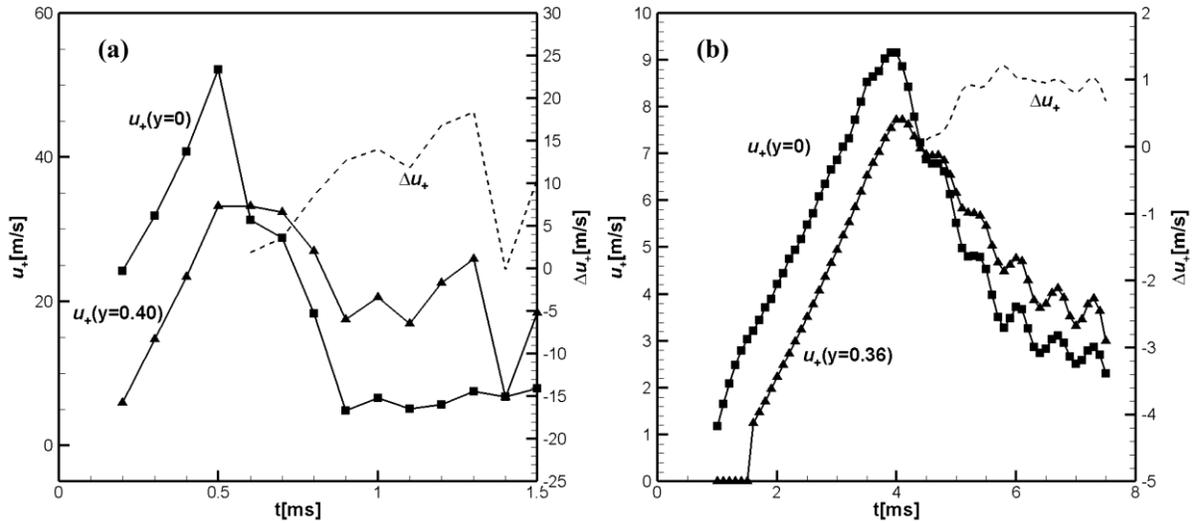

**Figure 5**: (a) Temporal evolution of the unburned flow velocities $u_+(y=0)$ and $u_+(y=0.4cm)$ in $H_2$/air; (b) the unburned flow velocities $u_+(y=0)$ and $u_+(y=0.36cm)$ in $CH_4$/air.

Figure 6 shows computed schlieren images and streamlines during tulip flame formation in (a) $H_2$/air and (b) $CH_4$/air mixtures in the channel with aspect ratio $\alpha=12$. The red dashed lines in the figures show the location of the unburned gas flow velocity profile measured at 0.5mm ahead of the flame front.

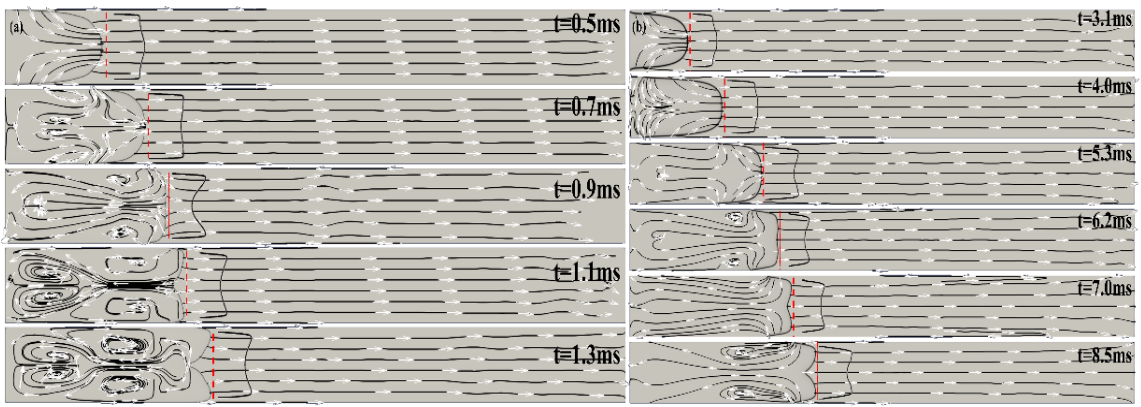

**Figure 6**. (a) Sequences of computed schlieren images, streamlines during tulip flame formation in H$_2$/air; and for Ch$_4$/air mixtures (b); the channel aspect ratio is $\alpha = 12$.

### 3.3. 2D semi-open channel

A semi-open channel can be considered as a limiting case of a channel with a very large aspect ratio, such that the pressure waves reflected from the opposite end have no time to reach and collide with the flame. In a semi-open channel, there is no reflected pressure waves and the pressure is almost constant. Without reflected pressure waves, which in a closed channel amplify the effect of the initial rarefaction wave and can initiate the RT instability leading to a distorted tulip flame, tulip flame formation in a semi-open channel takes longer time and proceeds smoothly.

Figure 7 shows the temporal evolution of the flame front velocities along the centerline and near the sidewall, the overpressure $P_+$ and the pressure growth rate $dP_+/dt$ calculated for the hydrogen/air (Fig.7a) and methane/air flames (Fig. 7b). The weak changes in the pressure growth rate and small oscillations of the flame front velocity seen in Fig. 7a are related to the moment when the flame skirt touches the sidewall, which causes a weak expansion wave propagating towards the channel centerline and reflecting from the sidewalls. They are much weaker for the methane/air flame in Fig. 7b. The temporal evolutions of the unburned gas velocities $u_+(y = 0)$, and $u_+(y = 0.25 cm)$ measured at 0.5mm ahead of the flame front are shown in Fig. 8(a, b) and it is seen that they also proceed more smoothly than for a channel with closed ends.

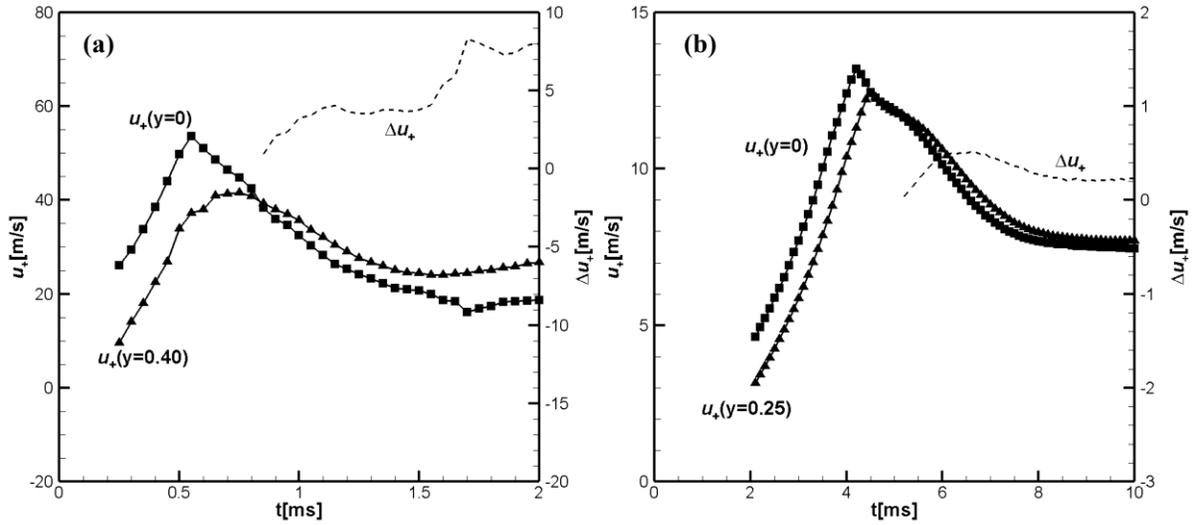

**Figure 8**: (a) Temporal evolution of the unburned flow velocities $u_+(y=0)$ and $u_+(y=0.4cm)$ in H$_2$/air mixture; (b) the unburned flow velocities $u_+(y=0)$ and $u_+(y=0.3cm)$ in CH$_4$/air mixture.

Figure 9 show sequences of computed schlieren images with streamlines during tulip flame formation in a semi-open tube. It can be seen that in both cases, for the fast hydrogen/air flame and for the slow methane/air flame, the tulip flame formation in a semi-open channel takes longer times than in a closed channel, but it is much faster for the hydrogen/air flame than for the methane/air flame. The red dashed lines in the figures show the location of the unburned gas flow velocity profile at 0.5mm ahead of the flame front.

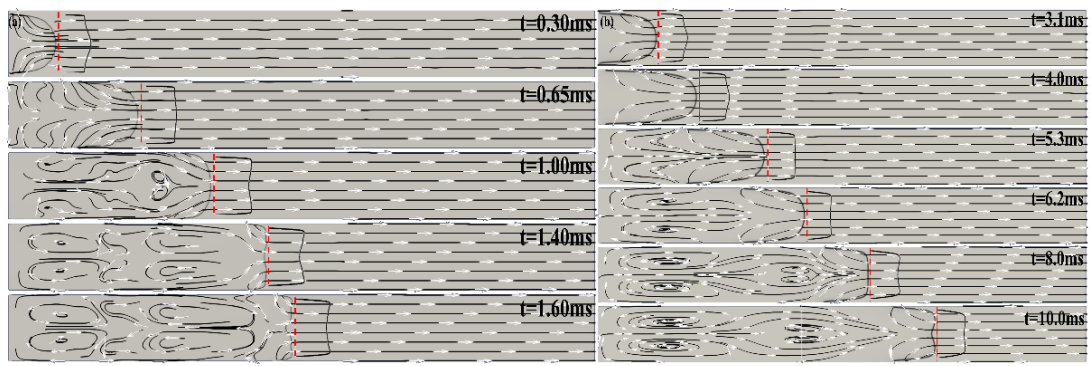

**Figure 9**. Sequences of computed schlieren images, streamlines during tulip flame formation in a semi-open tube; (a) hydrogen-air flame, (b) methane-air flame.

## 4. Summary and Discussions

It is found that 2D simulations provide a good qualitative agreement with experimentally observed flame dynamics, development of tulip and distorted tulip flames in the highly reactive hydrogen/air and in the low reactive methane/air flames, in particular the dependence on the channel aspect ratio [30, 47, 48]. It is shown that the laminar flame velocity strongly affects flame dynamics in the early stages of flame propagation, in agreement with numerical simulations by Deng et al. [9], which used a one-step chemical model. The effect of reaction order was studied by Qi et al. [10]. The present study shows that the reaction order has little effect on the formation of tulip flames, but may be important in later stages during the formation of distorted flames.

One of the important conclusions of this study is that laminar flame velocity and flame acceleration are the main factors responsible for the formation of tulip flame. In particular, the ratio of the tulip flame formation times for $H_2$/air and for $CH_4$/air mixtures is approximately equal to the ratio of their laminar flame velocities, $\tau_{tulip}(H_2)/\tau_{tulip}(CH_4) \approx U_{fL}(H_2)/U_{fL}(CH_4) \approx 6.3$.

Clanet and Searby [19] identified four characteristic times of the earlier stages of flame propagation in a tube: the time of transition from a spherical flame to a finger shape flame $\tau_{spher}$; the time when the flame skirt of the finger shape flame touches the sidewall $\tau_{wall}$; the time when flame inversion is initiated and the plane flame is formed $\tau_{inv}$; the time of tulip flame formation $\tau_{tulip}$. It should be emphasized that the hydrodynamic processes in 3D flames proceed considerably faster than those in 2D flames [26, 48]. Therefore, the characteristic times $\tau_{inv}$ and $\tau_{tulip}$ are longer for 2D flames than for 3D flames [26]. This explains the discrepancy between the theoretical prediction of the characteristic times $\tau_{inv}$ and $\tau_{inv}$ obtained using a simple 2D model, and the experimental times $\tau_{inv}$ and $\tau_{tulip}$ measured in [24, 49].

Using a geometric model by Clanet and Searby [19], we can write the equation for the flame tip velocity during the finger flame phase as

$$\frac{dX_{tip}}{dt} = 4\Theta \frac{U_{fL}}{D} X_{tip}, \qquad (12)$$

where $X_{tip}$ is the coordinate of the finger flame tip, $\Theta = \rho_u / \rho_b$ is the expansion factor, $D$ is the channel width, and $U_{fL}$ is the laminar flame velocity. Differentiating equation (12) with respect to time gives the dependence of flame acceleration on the laminar flame velocity

$$d^2 X_{tip} / dt^2 = (4\Theta U_{fL} / D)^2 X_{tip}. \qquad (13)$$

Equation (13) shows that the flame acceleration increases significantly as the flame velocity $U_{fL}$ increases. Since the amplitude of the pressure waves generated by the accelerating piston (flame front) is greater at higher piston accelerations, this means that a higher velocity flame generates stronger pressure waves than a lower velocity flame. It should be noted that in the phase of flame deceleration, when the flame skirt begins to touch the sidewalls of the tube, the main part of the finger flame skirt is almost parallel to the sidewalls of the tube, and it is larger for higher velocity flames. This also results in a large difference in the deceleration values of a high-speed flame compared to the deceleration of a lower-speed flame. For example, the numerical values of the flame accelerations ($a_+$) and deceleration ($a_-$) for the 2D channel considered in Sec. 3.1 are as follows: for H$_2$/air flame $a_+ \simeq 1.5 \cdot 10^5 m/s^2$ and $a_- \simeq -3 \cdot 10^5 m/s^2$; for CH$_4$/air flame $a_+ \simeq 3 \cdot 10^3 m/s^2$ and $a_- \simeq -6 \cdot 10^3 m/s^2$.

In recent paper by Lei et al. [50] it was assumed that effect of self-induced turbulence of burning velocity, and the flame wrinkling induced by DL instability is the main mechanism of tulip flame formation: "*Darrieus-Landau instability and thermal-diffusive (TD) instabilities are* *the main reason for the formation of tulip flames, and the DL instability played a dominant role*". However, no evidence has been provided that this is actually the case. The fact that tulip flame formation occurs faster than the characteristic time of the DL instability was established in numerical simulations long time ago by Gonzalez et al. [51] and Dunn-Rankin and Sawyer [52].

Here we show that the tulip flame formation due to the rarefaction wave takes much shorter time than the characteristic times of the Darrieus-Landau (DL) and thermal-diffusive (TD) instabilities. The time of the DL instability development can be estimated as $\tau_{DL} = 1/\sigma_{DL}$, where $\sigma_{DL} = kU_{fL}\frac{\Theta}{\Theta+1}\left(\sqrt{\Theta+1-1/\Theta}-1\right) \approx kU_{fL}\sqrt{\Theta}$ is the increment of the DL instability, $U_{fL}$ is the laminar flame speed, $\Theta = \rho_u/\rho_b$ is expansion coefficient, the ratio of unburned to burned gas densities, $k$ is the wave number. The time of establishing the reverse flow by a rarefaction wave can be estimated as

$$\tau_{RW} \approx D/a_s, \tag{14}$$

where $a_s$ is a speed of sound. Considering that the perturbation wavelength of the DL instability, which could be responsible for the flame front inversion, is about the channel width, $\lambda \sim D$, we obtain

$$\tau_{RW}/\tau_{DL} \approx \pi\sqrt{\Theta}\left(\frac{U_{fL}}{a_s}\right) \ll 1. \tag{15}$$

Since the laminar flame velocity is much less than the sound velocity, $U_{fL}/a_s \sim 10^{-3}-10^{-4}$, the inequality (15) holds for all flammable gas mixtures. In the same way, assuming for simplicity a unity Lewis number, it can be shown that the ratio of the characteristic time of the rarefaction wave $\tau_{RW}$ to the characteristic time of the thermal-diffusive instability $\tau_{TD}$ for $\lambda \sim D$ is

$$\tau_{RW} / \tau_{TD} \approx \frac{L_f}{D}\left(\frac{U_{fL}}{a_s}\right) \ll 1 \tag{16}$$

This means that, in agreement with the experimental studies by Ponizy et al. [27], Darrieus-Landau or thermally diffusive flame front instabilities are not involved in flame front inversion and tulip flame formation.

A decelerating flame front can be distorted by Rayleigh-Taylor instability if small perturbation can grow significantly during the deceleration time $\Delta t$. This occurs if the increment of the RT instability, $\sigma_{RT} = \sqrt{Akg}$, is sufficiently large to satisfy the condition $\sigma_{RT}\Delta t \gg 1$. Here $A = (\rho_u - \rho_b)/(\rho_u + \rho_b) = (\Theta - 1)/(\Theta + 1) \approx 1$ is the Attwood number, $k = 2\pi/\lambda$ is the wave number.

A necessary condition for the development of RT instability is negative flame acceleration. Negative flame acceleration first occurs during the flame deceleration phase, caused by a decrease in flame front surface area, when lateral portions of the flame skirt contact and collapse on the channel sidewalls. Negative flame acceleration can also be caused by the flame collision with reflected pressure waves. In the first case, calculating the flame acceleration $g$ during the flame front inversion, we find that for perturbation with wavelength, $\lambda \sim D$, which could be a possible candidate for flame front inversion, $\sigma_{RT}\Delta t < 1$.

The effects of flame collisions with reflected pressure waves have been studied by Qian and Liberman [53]. It was shown that the flame collisions with reflected pressure waves play an important role in the formation of the tulip flame and its further evolution to a distorted tulip flame. In particular, flame collisions with pressure waves enhance the effect of the first rarefaction wave generated by the decelerating flame in the unburned gas when it touches the tube sidewalls. This explains the more pronounced tulip flame formation in closed tubes compared to a semi-open tube observed in experiments and numerical simulations. In the later stages, the flame collisions with the reflected pressure waves can result in a much greater flame deceleration to meet the condition $\sigma_{RT}\Delta t \gg 1$. It was shown, that considering a finite thickness of the flame front, $L_f$, the maximum growth rate of the RT instability corresponds to perturbation $kL_f \approx 0.3$ [54], so the wavelength of the fastest growing mode of the RT instability is

$$\lambda_{max} = \frac{2\pi}{0.3}L_F \approx 0.2 cm \qquad (17)$$

The value $\lambda_{max} \approx 0.2cm$ agrees well with the size of the bulges developed at the tulip flame front in numerical and experimental studies of the distorted tulips in a closed duct [24, 25], see schlieren images at t=1.3ms in Fig. 3(a).

Since the intensity of the rarefaction wave is characterized by the flame acceleration after the flame skirt touches the sidewall, the times of flame front inversion and tulip flame formation are determined by the magnitude of the flame acceleration, which in turn depends strongly on the laminar flame velocity. Therefore, another conclusion is that at sufficiently low laminar flame velocities, the intensity of the rarefaction wave may be so small that a tulip flame formation takes a relatively long time. In this case, possible phenomena that may prevent tulip flame formation at low laminar flame velocities are the buoyancy effect resulting in faster development of the upper or lower lip, or sufficient time for flame instabilities to develop. This conclusion is consistent with the evolution of $CH_4/H_2$/air flames at different inhibitor percentages and hence different laminar flame velocities shown in Fig. 3 and with the experimental study of methane/hydrogen/air flames by Zhang et al [55].

A faster flame produces a stronger rarefaction wave during the finger phase of flame propagation. Therefore, for a faster flame, the difference between the unburned gas velocity at the tube axis and the axial (along the tube) velocity closer to the sidewalls is greater than for a flame with a slower velocity. This explains the deeper shape of the tulip formed by faster flames than that formed by slower flames.

**Appendix A**

Figure A1 shows the convergence validated by modeling a one-dimensional laminar $H_2$/air flame at initial pressure $P_0 = 2bar$, which corresponds to the maximum pressure observed in a 6 cm tube during the formation of a tulip flame, using the fixed grid resolutions $\Delta x = 10\mu m, 20\mu m, 40\mu m$. It can be seen that the temperature and velocity profiles converge to those obtained for $\Delta x = 10\mu m$. The red line shows the results obtained using an AMR grid with a coarse mesh $\Delta x = 40\mu m$ refined twice. It is clear that the AMR grid results are very close to the fixed grid results. Since the maximum pressure during tulip flame formation for $CH_4$/air flame is smaller than that for $H_2$/air flame, the resolution sufficient to resolve $H_2$/air flame is also used to simulate $CH_4$/air flame.

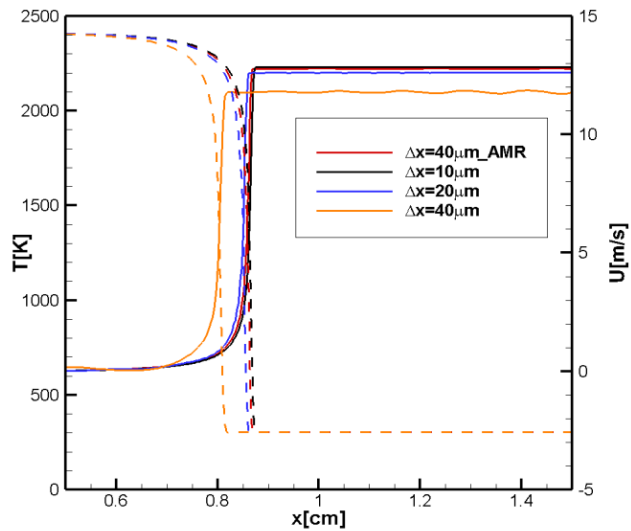

**Figure A1**. The temperature and flow velocity profiles of 1-D laminar flame of $H_2$/air at initial pressure $P_0 = 2bar$.

**Appendix B**

The pressure dependence of laminar flame velocity at given temperature $T_0 = 300K$ is

$$U_f / U_{f0} = (P/P_0)^{\frac{n}{2}-1}, \qquad (B1)$$

where $U_{f0}$ is the laminar flame velocity at $P_0 = 1\text{bar}$. On the other hand, the laminar flame velocity can be estimated as

$$U_f = D_{th} / L_f, \qquad (B2)$$

where $D_{th}$ is the thermal diffusion and $L_f$ is the laminar flame thickness. Combine two equations (B1) and (B2), we obtain the pressure dependence of laminar flame thickness as follows

$$L_f / L_{f0} = \left(\frac{P}{P_0}\right)^{-n/2}, \qquad (B3)$$

where $L_{f0}$ is the laminar flame thickness at $P = P_0$.

If we know $L_f$ at elevated pressures, we can obtain the reaction order $n$ by fitting Eq. (B.3) using least square method, as it is shown in Fig. (B1), where the empty triangles are calculated using Cantera and solid lines are plotted using Eq.(B.3) with $n=2$ for $H_2$/air flame, and $n=1.13$ for $CH_4$/air flame. It can be seen that the laminar flame thickness obtained using Eq. (B.3) agrees well with the data calculated by Cantera using detailed chemical models.

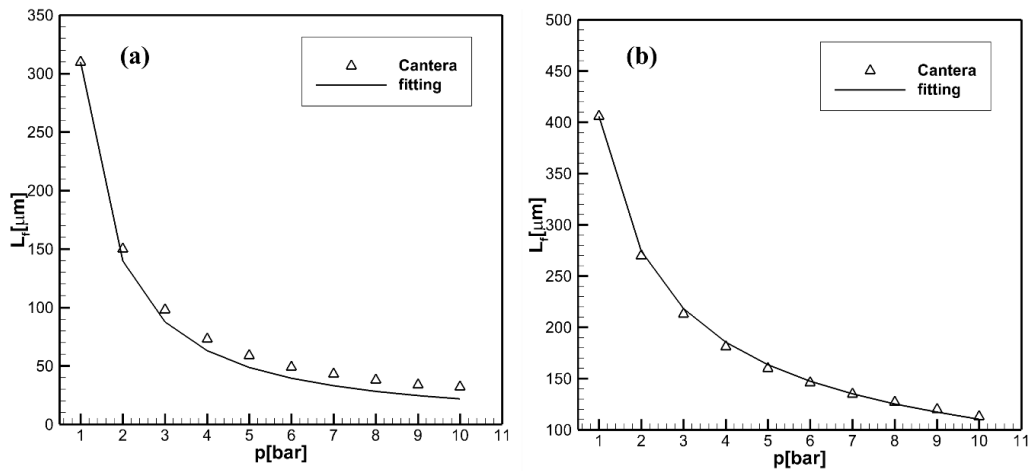

**Figure B1**(a, b). Laminar flame thickness of $H_2$/air and $CH_4$/air at elevated pressures. Empty triangles calculated by Canter; solid line: calculated using Eq.(B.3) with $n=2$ for $H_2$/air flame and $n=1.13$ for $CH_4$/air flame.

## Acknowledgements:

The authors benefitted from fruitful discussions with A. Brandenburg, P. Clavin and G. Sivashinsky. M. L. acknowledges the support for this work by the Olle Engkvists Stiftelse (Foundation) under grant No.232-0212, project No 31005258.